\documentclass{PoS}
\usepackage{amssymb,amsmath}
\usepackage[pdftex]{graphicx}

\def\dslash{{\partial\hskip-0.6em /}}

\def\Pslash{{P\hskip-0.5em /}}
\def\kslash{{k\hskip-0.45em /}}
\def\vslash{{v\hskip-0.5em /}}
\def\a{{\alpha}}
\def\g{{\gamma}}

\def\e{{\epsilon}}
\def\s{{\sigma}}
\def\D{{\Delta}}
\def\G{{\Gamma}}
\def\L{{\Lambda}}
\def\S{{\Sigma}}
\def\O{{\Omega}}
\def\mc#1{{\mathcal{#1}}}
\def\tr{{\mathrm{tr}}}

\title{New lessons from the nucleon mass, lattice QCD and heavy baryon chiral perturbation theory}

\ShortTitle{The nucleon mass and lattice QCD}

\author{\speaker{Andr\'{e} Walker-Loud}\\%\thanks{A footnote may follow.}\\
        Department of Physics, University of Maryland, College Park, MD 20742-4111, USA\\
        Department of Physics, College of William and Mary, Williamsburg, VA 23187-8795, USA\\
        E-mail: \email{walkloud@wm.edu}}

\abstract{I will review heavy baryon chiral perturbation theory for the nucleon delta degrees of freedom and then examine the recent dynamical lattice calculations of the nucleon mass from the BMW, ETM, JLQCD, LHP, MILC, NPLQCD, PACS-CS, QCDSF/UKQCD and RBC/UKQCD Collaborations. Performing the chiral extrapolations of these results, one finds remarkable agreement with the physical nucleon mass, from each lattice data set. However, a careful examination of the lattice data and the resulting extrapolation functions reveals some unexpected results, serving to highlight the significant challenges in performing chiral extrapolations of baryon quantities.  All the $N_f=2+1$ dynamical results can be quantitatively described by theoretically unmotivated fit function linear in the pion mass with $m_\pi \sim 750$--$190$~MeV.  When extrapolated to the physical point, the results are in striking agreement with the physical nucleon mass.  I will argue that knowledge of each lattice datum of the nucleon mass is required at the 1-2\% level, including all systematics, in order to conclusively determine if this is a bizarre conspiracy of lattice artifacts or rather a mysterious phenomenon of QCD.}

\FullConference{The XXVI International Symposium on Lattice Field Theory\\
		 July 14-19 2008\\
		 Williamsburg, Virginia, USA}

\begin{document}

\section{Introduction}
The nucleon mass provides an important benchmark calculation for lattice calculations.  Furthermore, the chiral expansion of the nucleon mass is now theoretically well understood, having been worked out to two-loop order in the framework of $SU(2)$ heavy baryon chiral perturbation theory~\cite{McGovern:1998tm}.  This theoretical tractability and the relative cleanliness with which it can be calculated from the lattice, make the nucleon mass a candidate for scale setting.  I would like to share with you some lessons I have recently learned about the nucleon mass, lattice QCD and chiral perturbation theory.  In particular, by examining all the recent (some preliminary) dynamical lattice calculations of the nucleon mass, from the BMW~\cite{BMW:latt08}, ETM~\cite{Alexandrou:2008tn}, JLQCD~\cite{Ohki:2008ff}, LHP~\cite{WalkerLoud:2008bp}, MILC~\cite{Bernard:2007ux}, NPLQCD~\cite{NPLQCD}, PACS-CS~\cite{Aoki:2008sm}, QCDSF/UKQCD~\cite{Gockeler:2007rx} and RBC/UKQCD~\cite{Yamazaki:2007mk} Collaborations, I will show that chiral extrapolations of the nucleon mass have some previously hidden surprises.

From the QCD trace anomaly, we know that the mass of all hadrons composed of valence light quarks, the $up$, $down$ and $strange$ flavors, are dominated by a quark mass independent term with chiral corrections which begin linear in the light quark masses,
\begin{equation}
	M_H = M_0^H + c_1^H m_q + \dots
\end{equation}
The exceptions to this rule are the eight pseudo-Nambu-Goldstone modes, $\phi = \{ \pi, K, \eta \}$, whose masses would vanish in the chiral limit, and otherwise scale as
\begin{equation}
	m_\phi^2 = B (m_{q_1} + m_{q_2}) + \dots
\end{equation}
The dots represent corrections to the masses which include terms polynomial in the quark masses as well as the much hunted for chiral logarithms and other chiral non-analytic functions.  The unambiguous discovery of predicted chiral non-analytic behavior in hadron observables is held by many to be necessary and un-refutable evidence that our lattice calculations have entered the \textit{chiral regime}, where the quark masses are light enough such that observables can be described by chiral perturbation theory ($\chi$PT)~\cite{Gasser:1983yg} and systematically extrapolated to the physical quark mass limit.

After reviewing heavy baryon $\chi$PT, I will examine the nucleon mass results from the above mentioned Collaborations and show they exhibit striking evidence of chiral non-analytic behavior.  However, the observed quark mass dependence is not predicted by any theoretical understanding of QCD we currently have, as the nucleon mass results do not obviously scale as $M_N = M_0^N + c_1^N m_q + c_{3/2}^N m_q^{3/2} + \dots$ as predicted, but rather as $M_N = \a_0^N + \a_{1/2}^N \sqrt{m_q}$, for pion masses in the range $m_\pi \sim 156$--$750$~MeV.  Given the variety of lattice actions used, as well as lattice spacings, lattice volumes and scale setting procedures, this behavior appears to be a phenomenon of QCD and not a conspiracy of lattice artifacts.  I conclude by exploring the precision which will be necessary to rule out this theoretically unmotivated quark mass dependence or accept it as a new mystery of QCD.

%\newpage
%
%%
%%%	HBChPT
%%%%
%%%%%
\section{Heavy Baryon Chiral Perturbation Theory}
I will begin with a review of heavy baryon chiral perturbation theory which was first developed by Jenkins and Manohar~\cite{Jenkins:1990jv}.  It is an effective field theory of non-relativistic baryons interacting with soft pions and photons, which provides a systematic expansion of baryon observables about the chiral limit.  For simplicity I will focus the two flavor theory in the isospin limit.%
%FOOTNOTE
\footnote{For an earlier and more recent review of HB$\chi$PT, see Refs.~\cite{Bernard:1995dp}.} 

When constructing an effective field theory of nucleons (and other baryons) difficulties arise immediately in developing a consistent power counting scheme.  Including the leading interaction with the axial field, $A_\mu = \partial_\mu \vec{\pi}\cdot \vec{\tau} / (\sqrt{2} f) +\dots$, the nucleon Lagrangian is
\begin{equation}\label{eq:LagRel}
\mathcal{L} = \bar{\psi}_N (i \dslash - M_0 ) \psi_N
	+g_A \bar{\psi}_N \g_\mu \g_5 A^\mu \psi_N
	+\dots
\end{equation}
leading to the equation of motion
\begin{equation}
	i \dslash\, \psi_N = M_0 \psi_N\, ,
\end{equation}
In all higher dimensional operators in the effective theory, we are free to replace derivatives acting on the nucleon field with its mass.  Phenomenologically, we know the chiral limit value of the nucleon mass is approximately the same size as the chiral symmetry breaking scale, $M_0 \sim \L_\chi$.  So how does one consistently count powers of derivatives, $(i\dslash / \L_\chi)^n \sim (M_0 / \L_\chi)^n \sim \mc{O}(1)$ when constructing the effective Lagrangian to a given order?

This is the question answered by Jenkins and Manohar.  When interacting with low energy pions and photons, the nucleon, in first approximation, can be treated as an infinitely heavy static field.  One begins by introducing a new set of nucleon fields related to the original nucleons by a phase, similar to that used to convert the Klein-Gordon equation to the Schr\"{o}dinger equation;
\begin{equation}
	N_v(x) = \frac{1+\vslash}{2}e^{iM_0 v\cdot x} \psi_N(x)\, .
\end{equation}
Here, $v_\mu$ is the four-velocity of the nucleon field.  In the rest frame, $v_\mu = (1,\mathbf{0})$ and the projector $(1+\vslash)/2 \rightarrow	(1+\g_0)/2$ becomes the non-relativistic projector.  The Lagrangian then takes the form
\begin{equation}\label{eq:HBLagLO}
\mc{L} = %\sum_v 
	\bar{N}_v iv\cdot \partial N_v
	+2 g_A \bar{N}_v\, S \cdot A N_v
	+2\a_M \bar{N}_v\, N_v \tr (\mc{M}_+)
	+\mc{O}\left(\L_\chi^{-1}\right)
	+\mc{O}\left(M_0^{-1}\right)\, .
\end{equation}
The mass term, $\bar{\psi}_N\, M_0 \psi_N$ is now absent from the Lagrangian, replaced with an infinite tower of $\mc{O}(M_0^{-n})$ suppressed operators, leading to a dual expansion.  In Eq.~\eqref{eq:HBLagLO}, I have additionally included the leading quark mass dependent term with the spurion field
\begin{align}
	&\mc{M}_+ = \frac{1}{2}\left( \xi m_Q^\dagger \xi + \xi^\dagger m_Q \xi^\dagger \right),&
	&\textrm{with}&
	&\xi^2 = \S = \textrm{exp}\left( \sqrt{2}i \vec{\pi}\cdot \vec{\tau} / f \right)&
\end{align}
and the quark mass matrix is $m_Q = \textrm{diag}(\hat{m}, \hat{m})$ with $2\hat{m} = m_u+m_d$.
The spin operator satisfies $S^2 N_v = -3/4 N_v$ and $v\cdot S=0$.
With this phase redefinition, it is convenient to decompose the nucleon momentum as
\begin{equation}\label{eq:P_nuc}
	P_\mu = M_0 v_\mu + k_\mu\, ,
\end{equation}
where $k_\mu$ is a residual soft momentum.  Derivatives acting on the nucleon field now bring down powers of this soft momentum, $\partial_\mu N_v = i k_\mu N_v$, 
which vanishes for on-shell nucleons at rest.  We see that the theory has a consistent power counting provided
$k^2 \sim m_\pi^2 \ll \L_\chi^2$.

\begin{figure}[t]
\center
\begin{tabular}{ccc}
\includegraphics[width=0.15\textwidth]{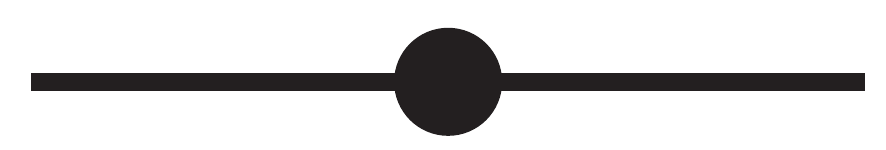}
&
\includegraphics[width=0.22\textwidth]{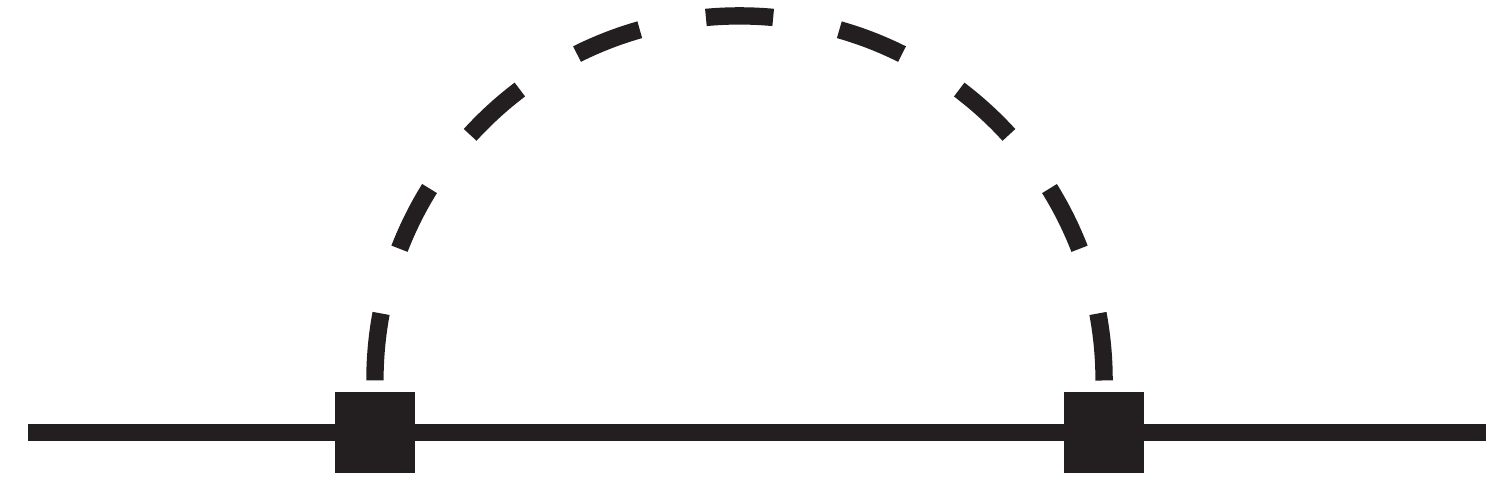}
&
\includegraphics[width=0.22\textwidth]{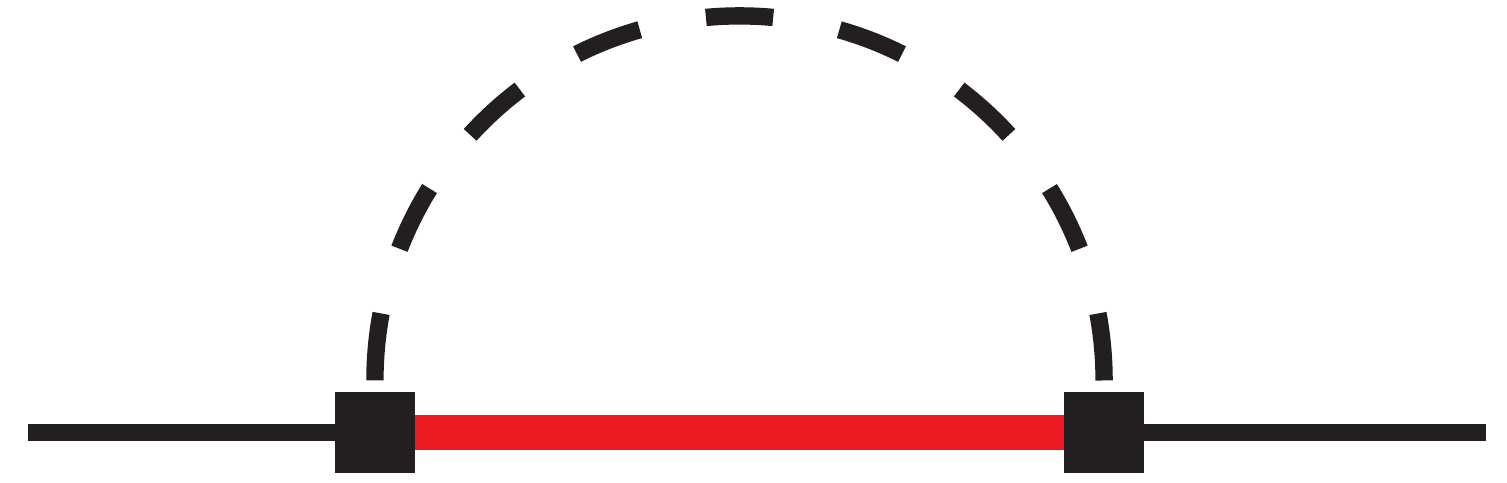}
\\
$(a)$ & $(b)$ & $(c)$
\end{tabular}
\caption{\label{fig:MN_LONLO} \textit{The leading order $(a)$ and next-to-leading order $(b)$ and $(c)$ contributions to the nucleon mass.}}
\end{figure}

One can also explicitly include the delta degrees of freedom in the theory.  However, this introduces a new mass parameter in the theory; the nucleon-delta mass splitting in the chiral limit
\begin{equation}
	\D_0 \equiv M_\D - M_N |_{m_q=0}\, .
\end{equation}
Phenomenologically, we know $\D \sim 290$~MeV.  In all present day lattice calculations, the pion mass is heavier than in nature, and so it is prudent to treat $\D_0 \sim m_\pi$ in the power counting.  One can package the delta fields in a totally symmetric flavor tensor and also utilize the Rarita-Schwinger representation for the spin.  This leads to the Lagrangian%
%FOOTNOTE
\footnote{The velocity subscript on the baryon fields is implicit here and following.} 
\begin{align}\label{eq:LagwithDelta}
\mc{L} =&\ \bar{N}\, i v\cdot D N + 2\a_M \bar{N} N \tr (\mc{M}_+) 
	- \bar{T}^\mu [iv\cdot D - \D_0 ] T_\mu - 2\g_M \bar{T}^\mu T_\mu \tr (\mc{M}_+ )
\nonumber\\&
	+2g_A \bar{N}\, S\cdot A N
	+2g_{\D\D}\bar{T}^\mu S\cdot A T_\mu
	+g_{\D N} \left( \bar{T}^\mu A_\mu N + \bar{N}\, A^\mu T_\mu \right)\, .
\end{align}
The axial couplings, $g_A$, $g_{\D N}$ and $g_{\D\D}$ are phenomenologically very important and these axial  interactions give rise to the leading non-analytic quark mass dependence of the baryon masses.  For example, the nucleon mass to leading loop order, depicted in Figure~\ref{fig:MN_LONLO} is
\begin{equation}\label{eq:MNnlo}
M_N = M_0 -2\a_M(\mu)m_\pi^2
	-\frac{3\pi g_A^2}{(4\pi f_\pi)^2} m_\pi^3
	-\frac{8g_{\D N}^2}{3(4\pi f_\pi)^2}\mc{F}(m_\pi,\D_0,\mu)
\end{equation}
where I have defined the standard chiral log function as
\begingroup
\small
\begin{align}
\mc{F}(m,\D,\mu) =&\ 
%\nonumber\\&
	(\D^2 - m^2 +i\e)^{3/2}
	\ln \left( \frac{\D+\sqrt{\D^2 - m^2 + i\e}}{\D - \sqrt{\D^2 - m^2 + i\e}} \right)
	-\frac{3}{2}\D m^2 \ln \left( \frac{m^2}{\mu^2}\right)
	-\D^3 \ln \left(\frac{4\D^2}{m^2}\right)\, .
\end{align}
\endgroup
The first term in Eq.~\eqref{eq:MNnlo} is the chiral limit value of the nucleon mass while the second term, denoted as the LO nucleon mass correction, comes from the tree graph, Fig.~\ref{fig:MN_LONLO}$(a)$.  The third and fourth terms are the NLO corrections to the nucleon mass arising from the one loop graphs depicted in Fig.~\ref{fig:MN_LONLO}$(b)$ and $(c)$ respectively.
%%%%%%%%%%%%%		FIGURE		%%%%%%%%%%%%%%%
%%%%%%%%%%%%%		END			%%%%%%%%%%%%%%%
Additionally, the explicit inclusion of the delta degrees of freedom leads to an infinite renormalization of the LEC $\a_M$, as well as a finite renormalization of $\a_M$ and $M_0$.  Furthermore, because the mass parameter $\D_0$ is a chiral singlet, it leads to an \textit{a priori} unknown finite renormalization of all LECs in the theory~\cite{WalkerLoud:2004hf}, for example 
\begin{align}
&M_0 \longrightarrow M_0 + \frac{16g_{\D N}^2}{9} \frac{\D_0^3}{\L_\chi^2} + d_3^M \frac{\D_0^3}{\L_\chi^2},&
&\a_M(\mu) = \a_M 
	- 2g_{\D N}^2 \frac{\D_0}{\L_\chi^2} \ln \left(\frac{\mu^2}{\mu_0^2}\right)
	+ \frac{2g_{\D N}^2}{3} \frac{\D_0}{\L_\chi^2} 
	+d_1^\a \frac{\D_0}{\L_\chi^2},&
\nonumber\\
&g_A \longrightarrow g_A \left( 1 + d_1^{g_A} \D_0 / \L_\chi +\dots \right),&
&g_{\D N} \longrightarrow g_{\D N} \left( 1 + d_1^{g_{\D N}} \D_0 / \L_\chi +\dots \right)\, .&
\end{align}
As we do not have the ability to dial the nucleon-delta mass splitting $\D_0$, as we can the quark masses, it is prudent to not keep explicit track of this finite renormalization but rather absorb it in a redefinition of the LECs.  In this way one accounts for all finite polynomial dependence upon $\D_0$.

The price one pays for this formalism is that Lorentz invariance is not manifestly preserved.  However, the symmetry is straightforward to recover order-by-order in inverse powers of $M_0$.  The method is known as reparameterization invariance~\cite{Luke:1992cs}.%
%FOOTNOTE
\footnote{For an equivalent formulation see Ref.~\cite{Bernard:1992qa}.} 
Consider a small shift in the the nucleon momentum, Eq.~\eqref{eq:P_nuc}, such that
\begin{align}
	&v \longrightarrow v + \e / M_0,&
	&k \longrightarrow k - \e\, .&
\end{align}
The nucleon momentum is invariant under this shift.  Requiring the S-matrix elements of the theory to also maintain invariance under this reparameterization recovers the Lorentz invariance order-by-order in $1/M_0$.  One manifestation of this requirement is a constraint on the coefficients of certain operators in the effective Lagrangian.  Specifically, the coefficients of certain higher dimensional operators are exactly determined from a connection with lower dimensional operators.  For example, considering the kinetic operators for the nucleon, one finds
\begin{equation}
\mc{L} = \bar{N}\, iv\cdot \partial N 
	- \bar{N}\frac{\partial^2}{2M_0} N + c\ \bar{N}\frac{(v\cdot \partial)^2}{2M_0} N
	+\mc{O} \left( M_0^{-2} \right)
\end{equation}
The coefficient of the first recoil operator is exactly constrained to be $-1$ while the coefficient of the second recoil operator $(1/2M_0) \bar{N} (v\cdot \partial)^2 N$ is unconstrained.  However, there is still a freedom to make a field-redefinition to convert the Lagrangian into the following form~\cite{Manohar:1997qy},
\begin{align}
&\mc{L} = \bar{N}\, iv\cdot \partial N 
	- \bar{N}\frac{\partial_\perp^2}{2M_0} N&
&\textrm{with}&
&\partial_\perp^2 = \partial^2 - (v\cdot \partial)^2\, .&
\end{align}
This choice provides the familiar form of the non-relativistic propagator, provided the reciol operator is re-summed to all orders,
\begin{equation}\label{eq:GN}
\frac{i(1+\vslash)/2}{v\cdot k + i\e} \longrightarrow
	\frac{i(1+\vslash)/2}{v\cdot k - \frac{\vec{k}^2}{2M_0} +i\e}\, .
\end{equation}
Calculating the nucleon mass to next-to-next-to-leading order (NNLO) requires this new recoil operator as well as other fixed-coefficient operators related to the axial coupling and a host of operators with unconstrained coefficients.  The calculation is still a one-loop calculation, but with operator insertions on the internal nucleon (delta) lines.  With the explicit inclusion of the delta degrees of freedom, the nucleon mass is given at NNLO by~\cite{WalkerLoud:2004hf}
\begingroup
\small
\begin{align}
M_N &=\ M_0 - 2\a_M(\mu)m_\pi^2
	-\frac{3\pi g_A^2}{(4\pi f_\pi)^2}m_\pi^3
	-\frac{8g_{\D N}^2}{3(4\pi f_\pi)^2}\mc{F}(m_\pi,\D_0,\mu)
%\nonumber\\&
%	+\frac{8g_{\D N}^2 (\a_M(\mu) - \g_M(\mu))}{(4\pi f_\pi)^2}m_\pi^2 \mc{J}(m_\pi,\D_0,\mu)
\nonumber\\&
	+\frac{m_\pi^4}{(4\pi f_\pi)^2}\ln \left( \frac{m_\pi^2}{\mu^2} \right) \left[
		6\a_M(\mu) - \frac{3b_A(\mu)}{4\pi f_\pi} - \frac{\frac{27}{8}g_A^2+5g_{\D N}^2}{2M_0} \right]
	+\frac{8g_{\D N}^2 (\a_M(\mu) - \g_M(\mu))}{(4\pi f_\pi)^2}m_\pi^2 \mc{J}(m_\pi,\D_0,\mu)
\nonumber\\&
	+m_\pi^4 \left[ b_M(\mu) +\frac{8 g_{\D N}^2 \a_M(\mu)}{(4\pi f_\pi)^2}
		-\frac{9 g_{\D N}^2}{4M_0(4\pi f_\pi)^2}
		-\frac{45g_A^2}{32M_0 (4\pi f_\pi)^2} \right]\, ,
\end{align}
\endgroup
where the new chiral log function is defined as
\begingroup
\small
\begin{equation}
\mc{J}(m,\D,\mu) = 
	2\D\sqrt{\D^2 - m^2+i\e} \ln \left( \frac{\D+\sqrt{\D^2 - m^2 + i\e}}{\D - \sqrt{\D^2 - m^2 + i\e}} \right)
	+m^2 \ln \left( \frac{m^2}{\mu^2} \right)
	+2\D^2 \ln \left( \frac{4\D^2}{m^2} \right)\, .
\end{equation}
\endgroup
The formula for the nucleon mass is in fact known to next-to-next-to-next-to-leading order (NNNLO) in a theory without explicit deltas~\cite{McGovern:1998tm}.  The heavy baryon theory has also been extended to quenched $\chi$PT~\cite{Labrenz:1996jy} as well as partially quenched $\chi$PT~\cite{Chen:2001yi} in which the baryon masses have been determined to NNLO~\cite{WalkerLoud:2004hf}.  In the framework of heavy baryon $\chi$PT, the leading and sub-leading lattice spacing corrections to the baryon masses have also been determined for Wilson fermions~\cite{Beane:2003xv}, twisted mass fermions~\cite{WalkerLoud:2005bt}, domain-wall on Wilson~\cite{Tiburzi:2005vy}, domain-wall on staggered~\cite{Tiburzi:2005is}, domain-wall on anything~\cite{Chen:2007ug} and staggered on staggered fermions~\cite{Bailey:2007iq}.

\subsection{Scalar Form Factor of the Nucleon\label{sec:ff}}
\begin{figure}[t]
\center
\begin{tabular}{cc}
\includegraphics[width=0.7\textwidth]{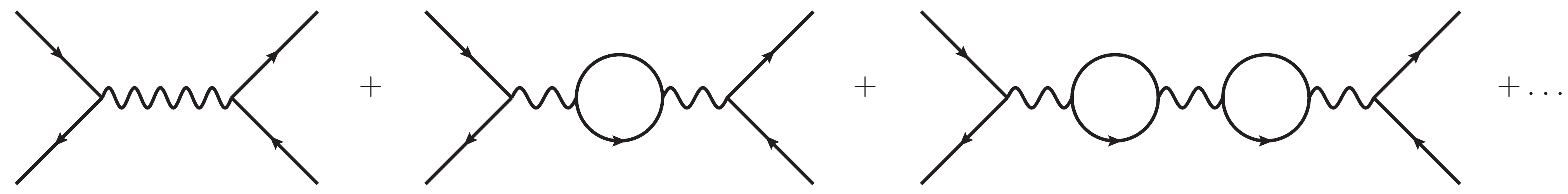}
&
\includegraphics[width=0.25\textwidth]{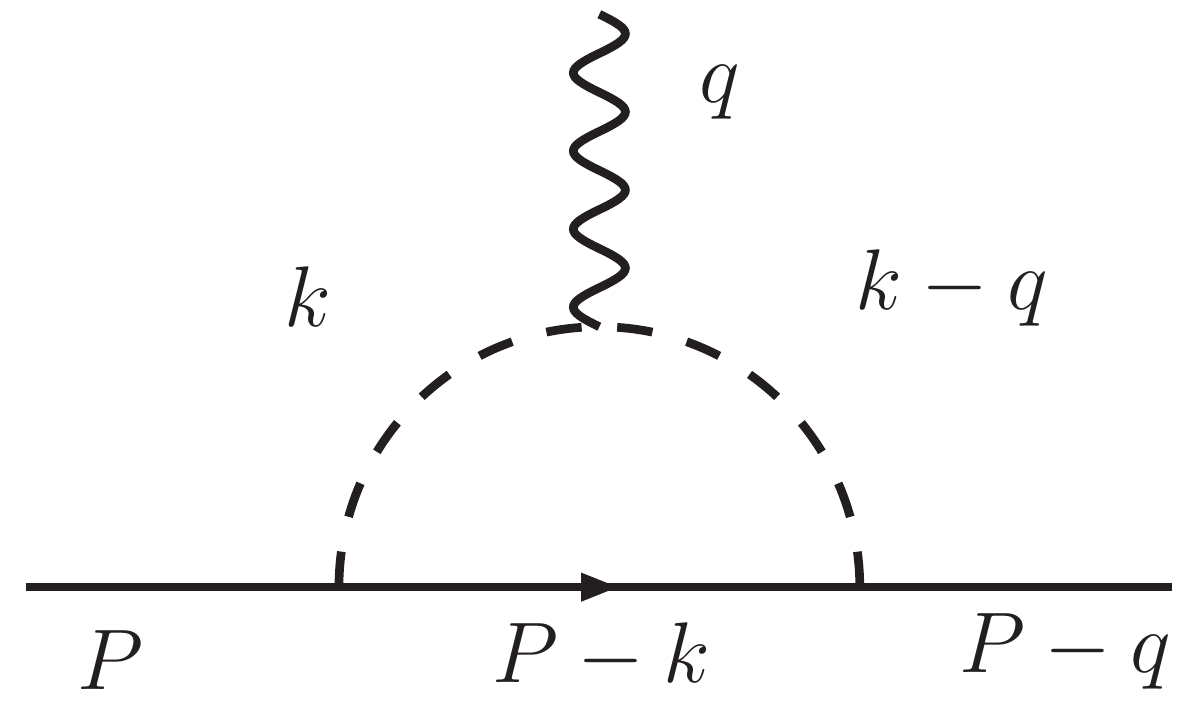}
\\
$(a)$ & $(b)$
%& + &
%\includegraphics[width=0.25\textwidth]{figures/epem1}
%& + &
%\includegraphics[width=0.25\textwidth]{figures/epem2}
\end{tabular}
\caption{\label{fig:epem} \textit{Perturbative graphs contributing to $e^+e^-$ scattering, $(a)$.  A one-loop contribution to the nucleon scalar form factor, $(b)$.}}
\end{figure}

The construction of heavy baryon $\chi$PT begins with the nucleon as infinitely heavy static fields.  Therefore, a naive use of the theory will lead to incorrect analytic structure in the formula for various observables.  Static quantities such as the nucleon mass, will be insensitive to these problems in the range of convergence of the theory.  However, dynamic quantities with external momentum insertions, such as pion-nucleon scattering, the scalar form factor, \textit{etc.} need to be analyzed with more care.  In some cases, for kinematic reasons, the power counting of the theory must be rearranged.  This is not a new phenomenon.  To provide a familiar example, consider $e^+e^-$ scattering near the Z-pole.  At any finite order in the perturbative calculation of the cross section, see Fig.~\ref{fig:epem}$(a)$, the sum will fail to reproduce the observed cross section.
%%%%%%%%%%%%%		FIGURE		%%%%%%%%%%%%%%%
%%%%%%%%%%%%%		END			%%%%%%%%%%%%%%%
Re-summing the self-energy of the Z will generated a width, leading to the correct form for the cross section,
\begin{equation}
\s_{e^+e^-}(s) \sim \frac{1}{(s-M_Z^2)^2}
	\longrightarrow \frac{1}{(s-M_Z^2)^2 +M_Z^2 \G_Z^2}\, .
\end{equation}

For a specific example relevant to heavy baryon $\chi$PT, I will consider the scalar form factor of the nucleon, defined by the matrix element,%
%FOOTNOTE
\footnote{The scalar form factor was first computed in heavy baryon $\chi$PT in Ref.~\cite{Bernard:1993nj} in the $SU(3)$ theory.} 
\begin{equation}
	\s(t=q^2) = \langle N(P) | m_q \bar{q} q |N(P-q) \rangle\, ,
\end{equation}
In Fig.~\ref{fig:epem}$(b)$, we depict one of the two one-loop diagrams contributing to this matrix element, the other having an internal delta state.  It is well known that the evaluation of this matrix element using the static nucleon propagator leads to an unphysical singularity.  So with hindsight in hand, I lets begin by re-summin the leading recoil corrections to the nucleon propagator, Eq.~\eqref{eq:GN}.
One then finds this loop contribution to the form factor
\begin{equation}
\s^{nlo}(q^2)	\sim \frac{g_A^2}{f^2} \int \frac{d^d k}{(2\pi)^d}
	\frac{\frac{1+\vslash}{2}\ h(S,k)}
		{\left( v\cdot k - \frac{\vec{k}^2}{2M_0}\right) ((k-q)^2 - m_\pi^2)(k^2 - m_\pi^2)}\, ,
\end{equation}
where $h(S, k)$ is a function of the spin operators and momenta.  The exact form of $h$ is not important for the manipulations I will make, as they will only modify the renormalization prescription.  First recall the coefficient of the operator $(1/2M_0) \bar{N} (v\cdot \partial)^2 N$ is not constrained as we are free to make a field redefinition to change it.  Therefore, let us make the change $\vec{k}^2 \rightarrow k^2$ in the nucleon propagator.  I will also factor out the $2M_0$ from the nucleon propagator, and with some malice aforethought, I will make a slight adjustment to $h(S,k)$, leaving us with
\begin{equation}
\s^{nlo}(q^2)	\sim \frac{g_A^2}{f^2} \int \frac{d^d k}{(2\pi)^d}
	\frac{[M_0\vslash - \kslash + M_0 ] h^\prime(S,k)}
		{( 2M_0 v\cdot k - k^2 ) ((k-q)^2 - m_\pi^2)(k^2 - m_\pi^2)}\, .
\end{equation}
For on-shell nucleons, $M_0\vslash = \Pslash$, such that the form factor takes the form
\begin{equation}\label{eq:ff3}
\s^{nlo}(q^2)	\sim \frac{g_A^2}{f^2} \int \frac{d^d k}{(2\pi)^d}
	\frac{\Pslash - \kslash + M_0}{M_0^2 - (P-k)^2}
	\frac{h^\prime(S,k)}
		{((k-q)^2 - m_\pi^2)(k^2 - m_\pi^2)}\, .
\end{equation}
Up to the numerator structure $h^\prime(S,k)$, this is exactly what we would have gotten by starting with the original relativistic Lagrangian, Eq.~\eqref{eq:LagRel}.  Becher and Leutwyler, in 1999, wrote a nice paper related to these issues~\cite{Becher:1999he}.  In addition to formally demonstrating the equivalence I have just discussed, they further showed how to regulate the integrals appearing with the re-summed nucleon propagator, as in Eq.~\eqref{eq:ff3}, in a manner which separates the chiral non-analytic pieces (the infrared pieces) from the polynomial in $m_q$ pieces (the regular pieces).  This has come to be known as \textit{covariant baryon $\chi$PT}.  Furthermore, the kinetic recoil corrections of the nucleon remove the unphysical singularity which occurs with a naive application of heavy baryon $\chi$PT, as I shall now demonstrate.  Evaluating the integral in Eq.~\eqref{eq:ff3} using the rules of Ref.~\cite{Becher:1999he}, one finds the scalar form factor is given at NLO by 
\begin{equation}
\s^{nlo}(t) = %-2\a_M m_\pi^2 
	\frac{3\pi g_A^2 m_\pi}{4(4\pi f_\pi)^2}(t-2m_\pi^2) \left[
		\sqrt{\frac{m_\pi^2}{t}} \ln \left( \frac{2+\sqrt{t/m_\pi^2}}{2-\sqrt{t/m_\pi^2}} \right)
		-\ln \left( 1 +\frac{m_\pi / (2 M_0)}{\sqrt{1-t/(4m_\pi^2)}} \right)
	\right]\, .
\end{equation}
Near the pion-production threshold, $t=4m_\pi^2$, there are singularities in both terms of this NLO contribution to $\s(t)$, which exactly cancel.  In a naive application of heavy baryon $\chi$PT, one would not have kept the second term, counting $m_\pi / (2M_0) \ll 1$ moving this term to NNLO and resulting in an unphysical singularity at $t=4m_\pi^2$ from the first term alone.  But as we have seen here, near this kinematic threshold, the power counting of the theory must be re-arranged to maintain a consistent and correct theory.  Near $t=4m_\pi^2$, one should count
\begin{equation}
	\frac{m_\pi}{2M_0} \sim \sqrt{1-\frac{t}{4m_\pi^2}}\, .
\end{equation}
Working away from the kinematic thresholds such as this, one is using the power counting 
\begin{equation}
	m_\pi \sim M_0\, .
\end{equation}
Furthermore, working away from these kinematic thresholds and not explicitly including the delta degrees of freedom is equivalent to choosing the power counting
\begin{equation}
	m_\pi \sim M_0 \ll M_\D - M_N\, .
\end{equation}

In summary, covariant baryon $\chi$PT is not a new effective field theory, but is heavy baryon $\chi$PT with a re-summed class of diagrams~\cite{Becher:1999he}.  The choice to use the original formulation of Jenkins and Manohar or to use the new covariant formalism of Becher and Leutwyler is a matter of taste.  As with all theories, the use of heavy baryon $\chi$PT, or the so derived chiral extrapolation formulae, requires consumer care; the naive use of either formalism may lead to inconsistencies.

%
%%
%%%	Comparing to LQCD results
%%%%
%%%%%
\section{Comparing with Lattice QCD Results}
%%%%%%%%%%%%%		FIGURE		%%%%%%%%%%%%%%%
\begin{figure}[t]
\center
\begin{tabular}{c}
\includegraphics[width=0.8\textwidth]{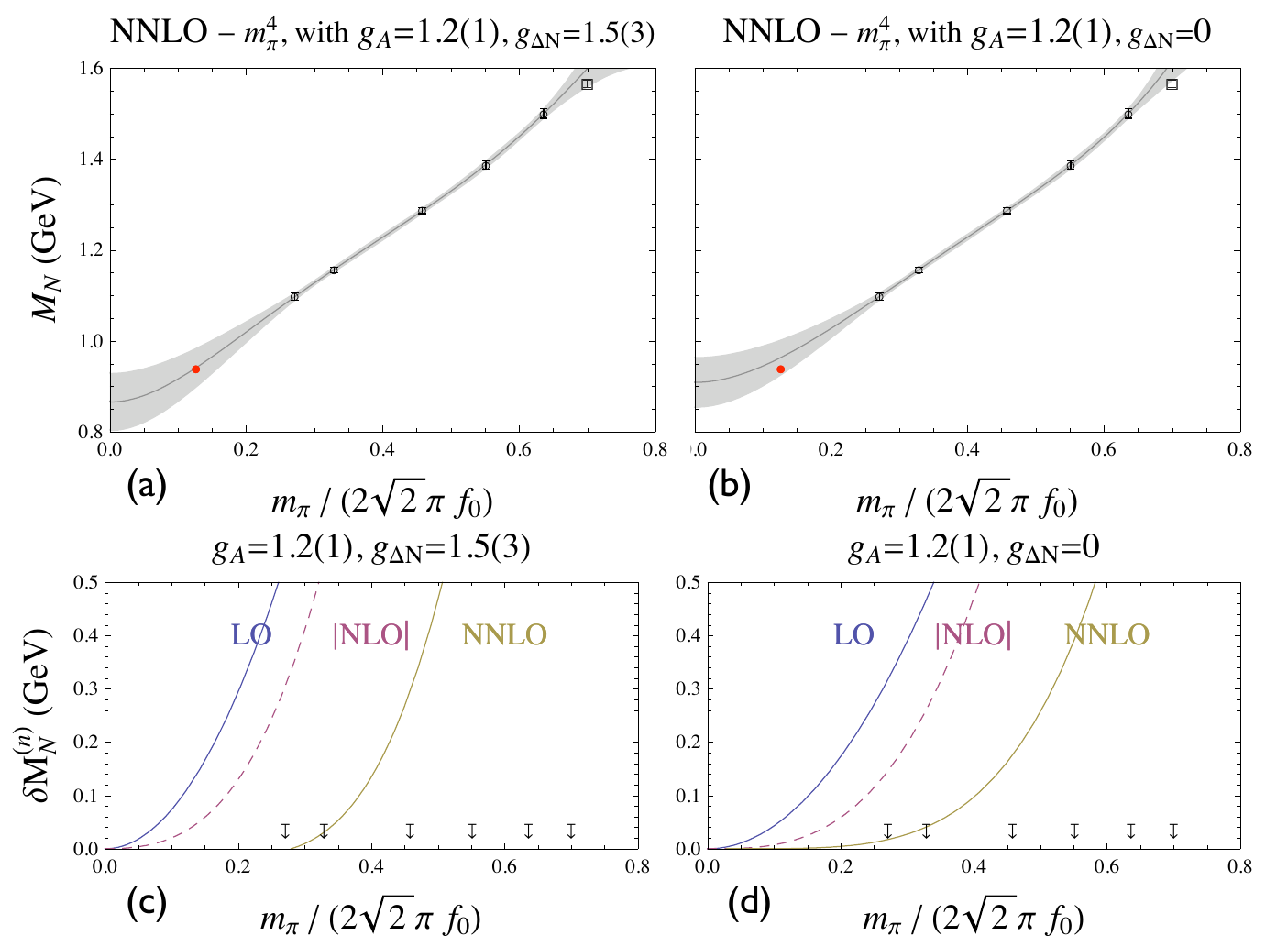}
%\includegraphics[width=0.45\textwidth]{figures/MNSU2NNLOFitErrPlot}
%&
%\includegraphics[width=0.45\textwidth]{figures/MNSU2NNLOnoDeltaFitErrPlot}
%\\
%$(a)$ & $(b)$
%\\
%\includegraphics[width=0.45\textwidth]{figures/MNLONLONNLOPlotandKey}
%&
%\includegraphics[width=0.45\textwidth]{figures/MNLONLONNLOnoDeltaPlotandKey}
%\\
%$(c)$ & $(d)$
\end{tabular}
\caption{\label{fig:LHPC_MN} \textit{Chiral extrapolation of LHPC nucleon mass results using the NNLO continuum formulae with $(a)$ and without $(b)$ explicit delta degrees of freedom.  In $(c)$ and $(d)$ we plot the corresponding contributions to the nucleon mass order by order.}}
\end{figure}
%%%%%%%%%%%%%		END			%%%%%%%%%%%%%%%
With a review of heavy baryon $\chi$PT in hand, I will now turn to comparing this formalism to all the recent dynamical lattice calculations of the nucleon mass.  I will begin with a review of the LHP results~\cite{WalkerLoud:2008bp}, with which I am the most familiar.  The LHP Collaboration performed a mixed-action calculation%
%FOOTNOTE
\footnote{The strange quark propagators and the majority of the light quark propagators for this calculation were performed by the NPLQCD Collaboration~\cite{Beane:2008dv}.} 
with domain-wall valence fermions on the coarse, $a\sim 0.125$~fm, $L\sim2.5$~fm Asqtad improved rooted staggered gauge ensembles with pion masses in the range $m_\pi \sim 290$--$750$~MeV.  The relevant mixed-action effective theory for this calculation is well established~\cite{Tiburzi:2005is,Chen:2007ug,Orginos:2007tw}.  However, there were too many unknown parameters in the mixed action formula to perform a successful chiral extrapolation.  Therefore, continuum $SU(2)$ heavy baryon $\chi$PT, including explicit delta degrees of freedom was used to perform the chiral extrapolations resulting in
\begin{equation}
	M_N = 954 \pm 42 \pm 20 \textrm{ MeV}\, .
\end{equation}
The first error is statistical while the second error is a systematic error resulting from varying the LECs which were fixed in the extrapolation, for specific details on the extrapolation analysis, see Ref.~\cite{WalkerLoud:2008bp}.  The resulting chiral extrapolation is depicted in Fig.~\ref{fig:LHPC_MN}$(a)$, plotted as a function of the pion mass over the approximate chiral symmetry breaking scale, $\L_\chi \sim 2\sqrt{2} \pi f_0$ where $f_0=121.9$~MeV~\cite{Colangelo:2003hf}.  A similar extrapolation was performed without deltas ($g_{\D N}=0$), resulting in the extrapolation depicted in Fig.~\ref{fig:LHPC_MN}$(b)$, which is qualitatively the same.  In Figs.~\ref{fig:LHPC_MN}$(c)$ and $(d)$ I plot the resulting order-by-order contribution to the nucleon mass from these two fits respectively.  In these plots, the \textit{arrows} denote the values of the pion mass at which the calculation was performed.  %One observes then that even at the lightest mass used in the calculation, the expansion is not great.  Not including the delta degrees of freedom, $(b)$ and $(d)$, appears to slightly improve the convergence of the expansion.  However, this should be taken with a large note of caution as excluding the deltas is assuming a power counting $m_\pi << M_\D - M_N$.

\begin{figure}[t]
\center
\begin{tabular}{c}
\includegraphics[width=0.8\textwidth]{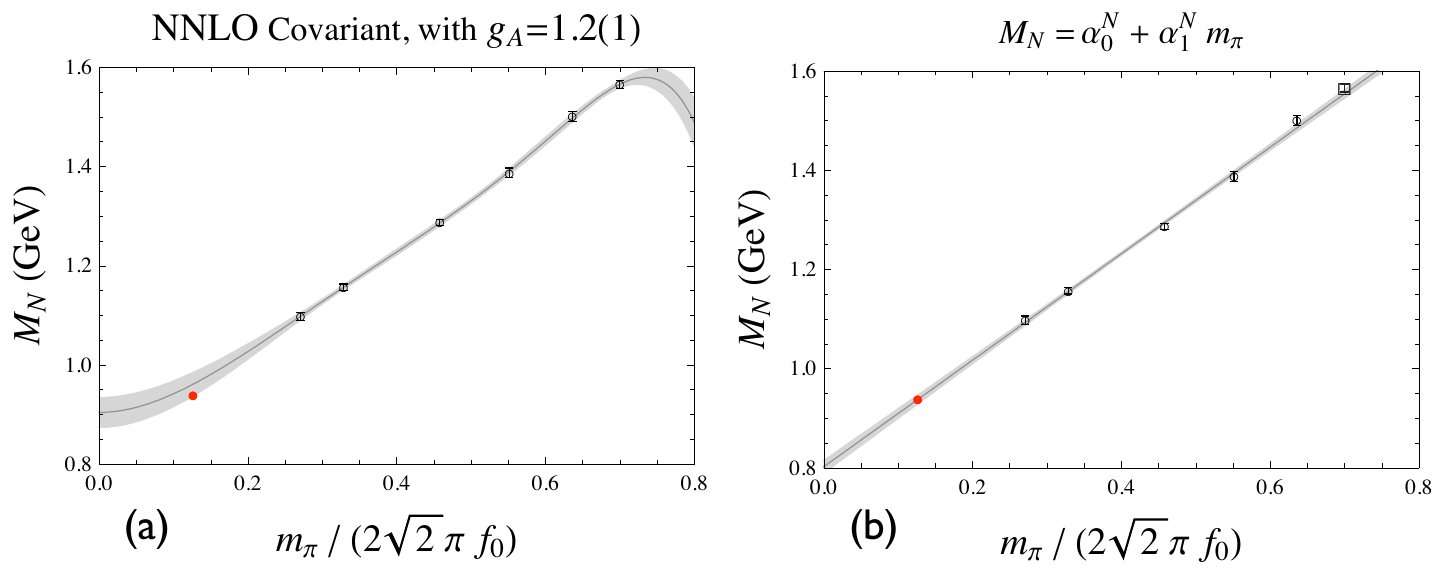}
%\includegraphics[width=0.42\textwidth]{figures/MNSU2NNLOCovFitErrPlot}
%&
%\includegraphics[width=0.42\textwidth]{figures/MNM0alphaFitErrPlot}
\end{tabular}
\caption{\label{fig:MN_cov} \textit{Nucleon mass extrapolation with full one-loop covariant formula, $(a)$.  A theoretically unmotivated extrapolation of $M_N$ linear in $m_\pi$, $(b)$.}}
\end{figure}

For comparison purposes, in this talk, I have also performed a chiral extrapolation using the full one-loop covariant baryon $\chi$PT formula, which can be found in Ref.~\cite{Procura:thesis}.  This extrapolation is depicted in Fig.~\ref{fig:MN_cov}$(a)$, which is again found to be qualitatively similar to the heavy baryon $\chi$PT fit.  The 68\% error band on the covariant fit is smaller than on the heavy baryon fits.  This is the result of more LECs being constrained, and not the result of the quality of the fit.
%%%%%%%%%%%%%		FIGURE		%%%%%%%%%%%%%%%
%%%%%%%%%%%%%		END			%%%%%%%%%%%%%%%

Despite the questionable convergence of heavy baryon $\chi$PT for these masses, the theory is consistent both with the lattice results and the physical point.  However, these fits all required phenomenological input.  Most notably, the axial couplings have to be fixed in these extrapolations, otherwise when left as free parameters, $g_A$ and $g_{\D N}$ float significantly away from their phenomenologically known values.  Furthermore, it is impossible not to notice the incredible linearity in the nucleon mass results as a function of $m_\pi$.  In fact, performing an extrapolation with a theoretically unmotivated linear fitting function (depicted in Fig.~\ref{fig:MN_cov}$(b)$), one finds
\begin{align}\label{eq:MN_ampi}
M_N &= \a_0^N + \a_1^N m_\pi
%\nonumber\\ &
	=938 \pm 9 \textrm{ MeV}\, ! \qquad\qquad \textrm{(with $\chi^2$/dof = 1.46)}
\end{align}
To highlight how \textit{wrong} this fit ansatz is, consider the pion-nucleon sigma term, which is the scalar form factor discussed in Sec.~\ref{sec:ff}, at zero momentum transfer,
\begin{equation}
\s_{\pi N} = \s(0) 
	= m_q \frac{\partial}{\partial m_q} M_N\, .
\end{equation}
Chiral symmetry dictates that $\s_{\pi N} / m_q \rightarrow \textrm{ const.}$ in the chiral limit, while for this fit function, Eq.~\eqref{eq:MN_ampi}, $\s_{\pi N} / m_q \rightarrow \infty$.
Still, one can not ignore the apparent phenomenological success of this fit in describing the LHP lattice results.  Is this a bizarre conspiracy of lattice artifacts or a physical phenomenon?  As the LHP calculation was performed at a single lattice spacing, and a single volume, one must examine the numerical calculations of other lattice groups to study this question.

%
%%
%%%					NF = 2+1
%%%%
%%%%%
\subsection{Comparing with $N_f=2+1$ calculations}
\begin{figure}[b]
\center
\begin{tabular}{c}
\includegraphics[width=0.75\textwidth]{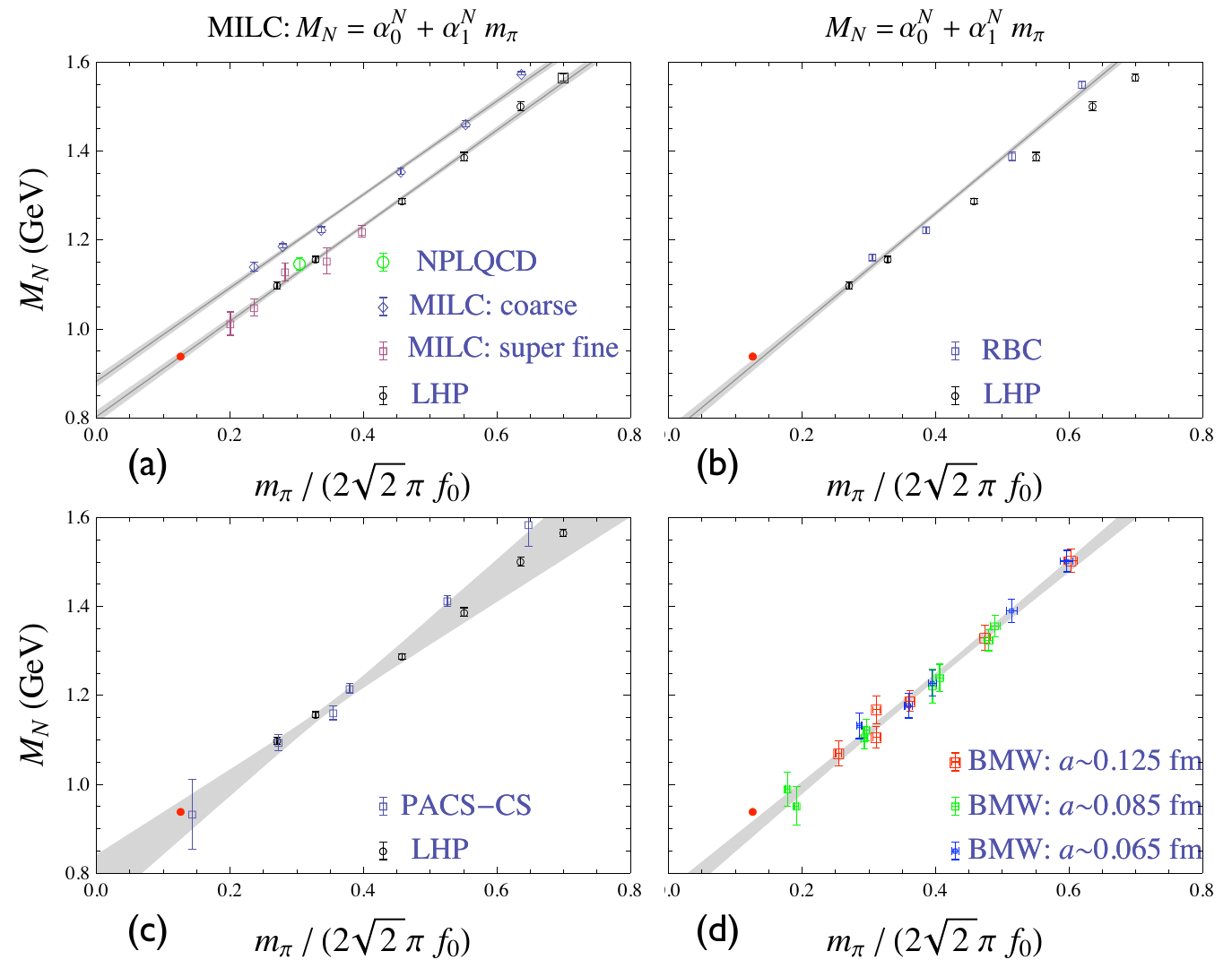}
%\includegraphics[width=0.42\textwidth]{figures/nucleonMassPlot}
%&
%\includegraphics[width=0.42\textwidth]{figures/RBCMNM0aFitErrPlot14}
%\\
%$(a)$ & $(b)$
%\\
%\includegraphics[width=0.42\textwidth]{figures/PACSMNM0aFitErrPlot14}
%&
%\includegraphics[width=0.42\textwidth]{figures/BMWMNM0aFitErrPlotALLLegend}
%\\
%$(c)$ & $(d)$
\end{tabular}
\caption{\label{fig:MNnf2+1} \textit{Nucleon mass results from all $N_f=2+1$ flavor lattice calculations.}}
\end{figure}

I will begin with an examination of the results from $N_f = 2+1$ calculations, those of the BMW Collaboration~\cite{BMW:latt08}, the MILC Collaboration~\cite{Bernard:2007ux}, the NPLQCD~\cite{NPLQCD}, the PACS-CS Collaboration~\cite{Aoki:2008sm} and the RBC/UKQCD Collaborations~\cite{Yamazaki:2007mk}.  In Table~\ref{tab:lattStat} I list relevant parameters from these lattices.  As can be seen, there is a wide range of parameters used as well as several independent methods of scale setting.  In Figure~\ref{fig:MNnf2+1}, I plot the resulting values of the nucleon mass from these collaborations, in all cases using the scale setting method of the given Collaborations to convert all results to GeV.  When it does not crowd the plot too much, I also add the LHP results as a reference.
%%%%%%%%%%%%%		TABLE		%%%%%%%%%%%%%%%
\begin{table}[t]
\center
{\footnotesize
\caption{\label{tab:lattStat} Details of the $N_f=2+1$ dynamical lattice calculations.}
\begin{tabular*}{0.95\textwidth}{@{\extracolsep{\fill}}ccccc}
\hline\hline
Collaboration & a[fm] & L[fm] & $m_\pi L \geq$ & scale setting \\
\hline
RBC/UKQCD & 0.114 & 2.74 & 4.56 & $M_{\O^-}$\\
PACS-CS & 0.091 & 2.90 & 2.29 & $r_0$\\
NPLQCD & 0.125, 0.09 & 2.5 & 3.7 & $r_1, r_0$ \\
MILC & 0.125, 0.09, 0.06 & 2.5, 3.0, 3.5 & 3.7 & $r_1,r_0$ \\
LHP & 0.125 & 2.5 & 3.7 & $r_1, r_0$ \\
BMW & 0.125, 0.085, 0.065 & 2.0, 3.0, 4.0 & 4 & $M_\Xi$, $M_{\O^-} \dots$\\
\hline\hline
\end{tabular*}}
\end{table}
%%%%%%%%%%%%%		END			%%%%%%%%%%%%%%%

In Fig.~\ref{fig:MNnf2+1}$(a)$, one observes the coarse ($a\sim 0.125$~fm) MILC results are systematically higher than the LHP results.  This is most likely an additive lattice spacing correction to the nucleon mass.  The super-fine ($a\sim 0.06$~fm) MILC results are very consistent with the LHP results with the lower error band coming from a fit to the LHP results alone.  The super-fine MILC results, which are preliminary, are consistent with this linear in $m_\pi$ extrapolation down to $m_\pi\sim 220$~MeV.  In Fig.~\ref{fig:MNnf2+1}$(a)$, I have also added a preliminary result from NPLQCD on the fine ($a\sim 0.09$~fm) MILC lattices with domain-wall valence fermions, which is also consistent with the LHP results.  This is suggestive that the discretization errors in the mixed-action domain-wall valence on MILC sea fermions are small.
%%%%%%%%%%%%%		FIGURE		%%%%%%%%%%%%%%%
%%%%%%%%%%%%%		END			%%%%%%%%%%%%%%%
In Fig.~\ref{fig:MNnf2+1}$(b)$ I have plotted the preliminary RBC/UKQCD results which are an update of those in Ref.~\cite{Yamazaki:2007mk}.  One observes the lighter two points are consistent with the LHP results while the heavier two points are inconsistent as measured with the statistical errors.  However, RBC/UKQCD use an independent method of scale setting, which may account for this discrepancy.  In Fig.~\ref{fig:MNnf2+1}$(c)$, I have plotted the recent results from PACS-CS~\cite{Aoki:2008sm}, which are also consistent with those of LHP.  The error band in Fig.~\ref{fig:MNnf2+1}$(c)$ is the result of fitting the lightest four PACS-CS results.  Although their lattice spacing is reasonably small, the lightest PACS-CS result, at the impressive $m_\pi\sim156$~MeV has a rather small value of $m_\pi L$ and is thus expected to have a significant volume correction, although perhaps not bigger than the current statistical error bar.  The BMW Collaboration has the most impressive set of results, plotted in Fig.~\ref{fig:MNnf2+1}$(d)$.  They have results at three different lattice spacings, and all points plotted have $m_\pi L \geq 4$, with the lightest pion mass $m_\pi\sim 190$~MeV.  Within the statistical errors, there are no discernible discretization errors.

Given the large number of independent results from the various collaborations, and with the three independent scale setting methods, we can take the trends of the nucleon mass results as a function of $m_\pi$ seriously.  The most striking feature is that all these $N_f=2+1$ results display this remarkably linear dependence upon the pion mass.  For the range of pion masses at which these calculations were performed, it would seem that QCD has conspired to produce this highly unexpected chiral non-analytic behavior of the nucleon mass, such that
$M_N \sim \a_0^N + \tilde{\alpha}_1^N \sqrt{m_q}$.
Before discussing what is needed to rule out this theoretically unmotivated chiral behavior in favor of that expected from heavy baryon $\chi$PT, I will examine the results of the $N_f=2$ flavor dynamical calculations.

%
%%
%%%				NF = 2
%%%%
%%%%%
\subsection{Comparing with $N_f=2$ calculations}
\begin{figure}[t]
\center
\begin{tabular}{c}
\includegraphics[width=\textwidth]{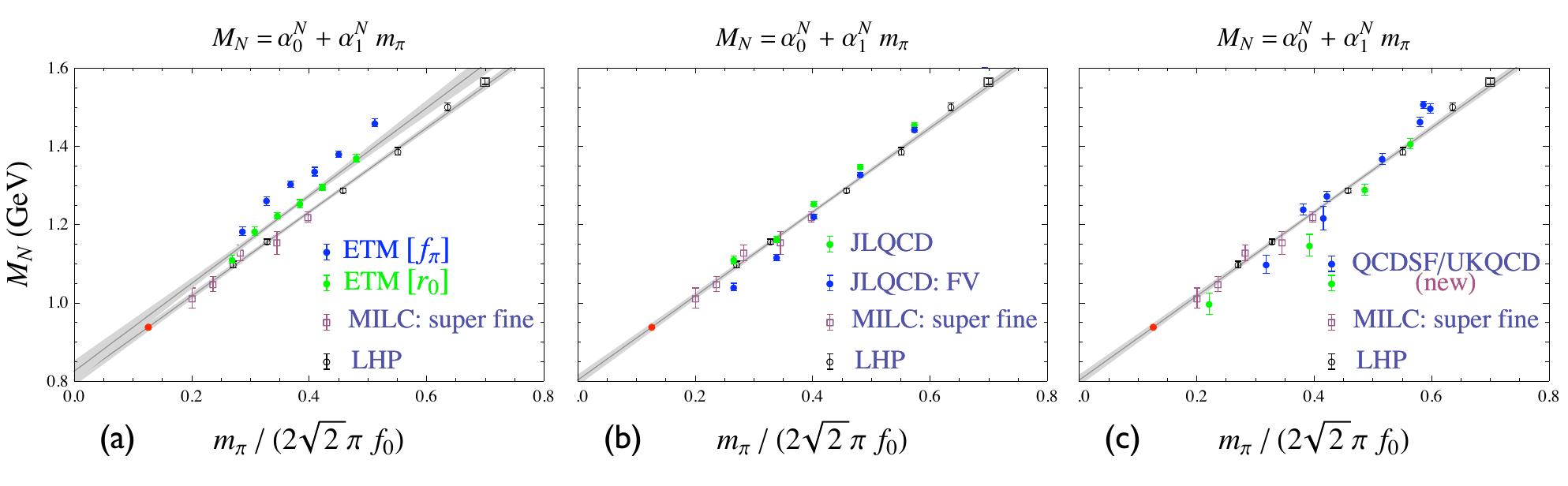}
%&
%\includegraphics[width=0.32\textwidth]{figures/ETMPlot}
%\\
%$(a)$ & $(b)$
%\\
%&
%\includegraphics[width=0.32\textwidth]{figures/JLQCDComparePlot}
%&
%\includegraphics[width=0.32\textwidth]{figures/QCDSFPlot}
%\\
%
%$(a)$ & $(b)$ &$(c)$
\end{tabular}
\caption{\label{fig:MNnf2} \textit{Nucleon mass results from all $N_f=2$ flavor lattice calculations.}}
\end{figure}

I now make a similar examination of the results from the $N_f=2$ calculations from the ETM~\cite{Alexandrou:2008tn}, JLQCD~\cite{Ohki:2008ff} and QCDSF/UKQCD~\cite{Gockeler:2007rx} Collaborations.  In Table~\ref{tab:lattStatNf2}, I collect pertinent parameters from the different lattices used in the calculations.
%%%%%%%%%%%%%		TABLE		%%%%%%%%%%%%%%%
\begin{table}[b]
\center
{\footnotesize
\caption{\label{tab:lattStatNf2} Details of the $N_f=2$ dynamical lattice calculations.}
\begin{tabular*}{0.95\textwidth}{@{\extracolsep{\fill}}ccccc}
\hline\hline
Collaboration & a[fm] & L[fm] & $m_\pi L \geq$ & scale setting \\
\hline
ETM & 0.083, 0.0655 & 2.10, 2.66 & 3.3 & $r_0$ \& $f_\pi$\\
JLQCD & 0.118 & 1.9 & 2.8 & $r_0$\\
QCDSF/UKQCD & 0.085--0.067 & 1.5--2.6 & 2.1 & $r_0$ \\
\hline\hline
\end{tabular*}}
\end{table}
%%%%%%%%%%%%%		END			%%%%%%%%%%%%%%%
In Fig.~\ref{fig:MNnf2}, I plot the resulting values of the nucleon mass, using the scale setting methods employed by the respective groups.  The $N_f=2$ calculations all have smaller volumes (and minimum values of $m_\pi L$) than the $N_f=2+1$ calculations.  There is also less independence in scale setting methods; all three groups use $r_0$~\cite{Sommer:1993ce} while ETM also uses $f_\pi$ to set the scale.  As compared to the $N_f=2+1$ calculations, the $N_f=2$ calculations are more varied with respect both to each other as well as with respect to the coarse LHP and super-fine MILC results which I am using as a reference.  With either scale setting method, the ETM results (continuum extrapolated), depicted in Fig.~\ref{fig:MNnf2}$(a)$, are systematically higher than the LHP/MILC results, with the results using $f_\pi$ to set the scale~\cite{Urbach:2007rt} being the most different.  The ETM results have been computed in smaller volumes (at smaller $m_\pi L$) and the predicted one-loop volume corrections~\cite{Beane:2004tw} will lower the results at the few percent level, bringing them in closer agreement with the LHP/MILC results.
%%%%%%%%%%%%%		FIGURE		%%%%%%%%%%%%%%%
%%%%%%%%%%%%%		END			%%%%%%%%%%%%%%%
The results of the JLQCD Collaboration ($a\sim 0.118$~fm clover fermions) are depicted in Fig.~\ref{fig:MNnf2}$(b)$.  The raw numbers are in approximate agreement with the LHP/MILC results.  However, the JLQCD results were calculated with smallish volumes.  They adjusted the results for predicted finite volume corrections, whcih I have listed as \textit{JLQCD: FV} in Fig.~\ref{fig:MNnf2}$(b)$.  One observes these numbers at the lightest two points are not in agreement with the LHP/MILC results.  Furthermore, they display definite curvature (non linearity in $m_\pi$ behavior).  The QCDSF/UKQCD results ($a\sim 0.085$--$0.067$~fm) are expected to have the largest finite volume corrections, for which these numbers have not been adjusted.  The results are from Ref.~\cite{Gockeler:2007rx} as well as preliminary new numbers.  The (new) preliminary numbers from QCDSF/UKQCD are not in agreement with the LHP/MILC results, and the lighter points are expected to have larger finite volume corrections, possibly as large as 5-10\%.  However, the results from Ref.~\cite{Gockeler:2007rx} are in agreement with the LHP/MILC numbers.

Although the $N_f=2$ nucleon mass results are in less agreement with the LHP/MILC results (as compared to all the $N_f=2+1$ results which are all in good agreement), aside from the JLQCD finite volume adjusted results, the $N_f=2$ results are also consistent with the theoretically unmotivated linear in $m_\pi$ behavior.
There appears to be overwhelming ``lattice phenomenological" evidence that this unexpected chiral non-analytic behavior is a real QCD phenomenon and not simply a bizarre coincidence of lattice artifacts.  A natural question to ask is what level of precision is required in our numerical nucleon mass results to rule out this possibility?  I will address this question in the next section.

%
%%
%%%	Challenges for Baryon Observables
%%%%
%%%%%
\section{Challenges with chiral extrapolations for baryon observables}
\begin{figure}[t]
\center
\begin{tabular}{c}
\includegraphics[width=0.8\textwidth]{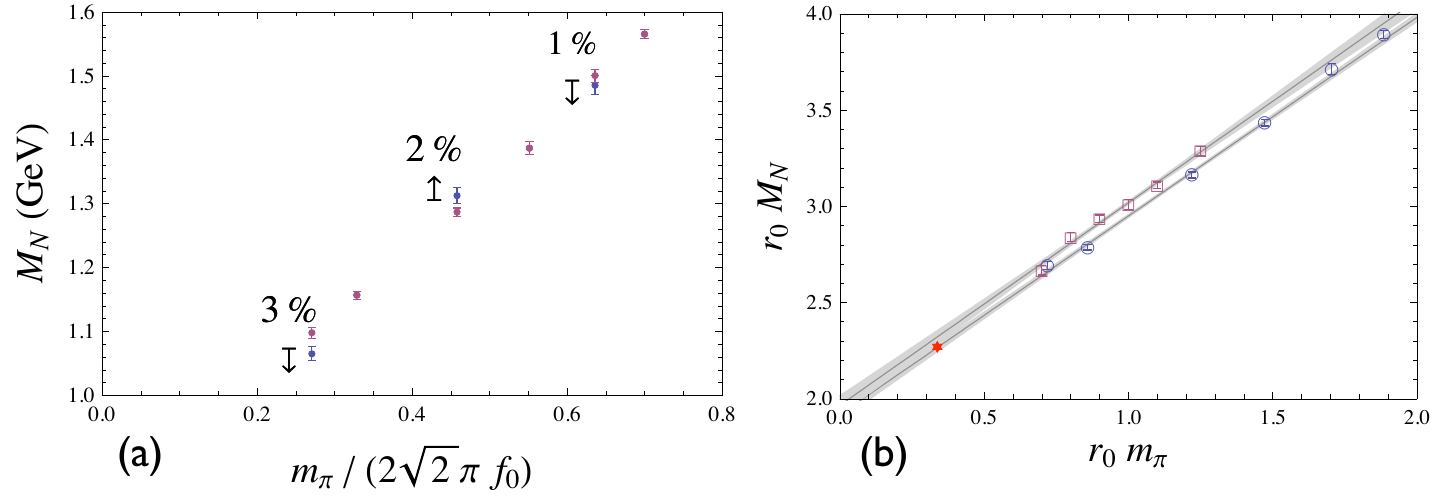}
%\includegraphics[width=0.42\textwidth]{figures/MN1percPlot}
%&
%\includegraphics[width=0.42\textwidth]{figures/MNlhpetmM0aFitErrPlot2}
%\\
%$(a)$ & $(b)$
\end{tabular}
\caption{\label{fig:MNr0} \textit{Curvature added to the LHP results, $(a)$.  ETM, and LHP nucleon mass results in $r_0$ units, $(b)$.}}
\end{figure}

Given the numerical state of affairs, in particular the nucleon mass results from the $N_f=2+1$ calculations, I can only make qualitative statements about the needed precision to rule out the straight-line fit in favor of the heavy baryon $\chi$PT analysis.  To quantify my qualitative statements, I will use a $\chi^2/dof \geq 2$ to rule out a particular fit.  I will continue to use the LHP results~\cite{WalkerLoud:2008bp} as my reference.  I will perform the following excercise; by hand I will lower the lightest mass results by 3\%, raise the third lightest point by 2\%, lower the fifth lightest point by 1\% and leave the second, fourth and heaviest points as they are.  In this way I am adding curvature to the results which can be accomodated with the unknown LECs in the heavy baryon $\chi$PT extrapolation formula, but which will rule out the linear in $m_\pi$ fit.  See Fig.~\ref{fig:MNr0}$(a)$.
%%%%%%%%%%%%%		FIGURE		%%%%%%%%%%%%%%%
%%%%%%%%%%%%%		END			%%%%%%%%%%%%%%%
One can then ask \textit{how small must the error bars be on the nucleon mass results to rule out the straight-line in $m_\pi$ fit}?  The answer is 1\% which returns a $\chi^2/dof = 2.08$.  With 2\% error bars, the $\chi^2/dof = 0.52$, in both cases with four degrees of freedom.  This means we would have to know each nucleon mass point at the 1\% level including all sources of systematic error!  A challenging endeavor to say the least.

At this level of precision, all systematic errors may come into play; the lattice spacing corrections (present in all but perfect lattice actions), finite volume corrections, heavy quark corrections (from not having the $c$, $b$ and $t$ dynamical quarks), scale setting errors, \textit{etc.}  From the analysis in this talk, it appears that scale setting may introduce one of the larger uncertainties.  All the $N_f=2+1$ calculations are in good agreement with each other, and have used three independent scale setting methods.  However, the ETM results have been determined with two independent scale setting methods, the use of $r_0$ and also the use of $f_\pi$.  From Fig.~\ref{fig:MNnf2}$(b)$, one can see a striking difference between the two methods which should in principle agree.  If one instead plots the LHP and ETM results in $r_0$ units, the disagreement between the two sets of results becomes even less pronounced, although still with a slight systematic shift.  See Fig.~\ref{fig:MNr0}$(b)$.
Given the lightest quark mass results of the BMW and MILC Collaborations, it seems evident that the resolution of this puzzle does not lie with lighter quark masses, at least those which are light but still heavier than the physical pion mass.  Collectively, we will need to address all the known lattice artifacts with numerical calculations at finer lattice spacings, larger volumes, and perhaps more care in addressing the errors associated with scale setting and comparing scale setting methods.

%
%%
%%%	Concluding remarks
%%%%
%%%%%
\section{Conclusions}
The nucleon mass has often been thought of as an important and clean benchmark observable with which to calibrate ones lattice calculations, and a candidate quantity for scale setting.  By clean, I mean relatively free of systematic uncertainties.  Little did we know that (lattice) QCD has conspired to make the nucleon mass far more mysterious and wrought with uncertainty than previously thought.  For years, we as a lattice community have been searching for the tell-tail signs of the chiral logs (and other predicted chiral non-analytic functions).  A close examination of calculations of the nucleon mass reveals that the results from all dynamical $N_f=2+1$ and some of the $N_f=2$ lattice calculations can be very well described by the chiral non-analytic (but unexpected) function
\begin{equation*}
	M_N =\a_0^N + \a_1^N m_\pi\, .
\end{equation*}
Perhaps more shocking, extrapolations of the lattice results to the physical pion mass produce a number in striking agreement with the physical nucleon mass; from the LHP results one finds $M_N = 938\pm9$~MeV.  Given the large amount of independent lattice actions, lattice spacings, lattice volumes and scale setting methods, this appears not to be a bizarre conspiracy of lattice artifacts but rather a phenomenon from QCD.  Resolution of this puzzle (ruling out this linear in $m_\pi$ function in favor of the heavy baryon $\chi$PT formula) may require nucleon mass results with uncertainties at the sub 1\% level, including all systematic errors.  

One thing is now clear, the nucleon mass is a bad observable to use to hunt for predicted chiral non-analytic behavior.  A determination of the phenomenologically interesting~\cite{Kaplan:2000hh} pion-nucleon sigma term, $\s_{\pi N}$, I predict will further require lattice calculational results on both sides of the physical point.  From the LHP numbers, one finds an extrapolation including explicit delta degrees of freedom results in $\s_{\pi N} \sim 84\pm17\pm20$~MeV.  The result without explicit deltas returns a value closer to the current consensus, $\s_{\pi N}\sim 42\pm14\pm9$~MeV.  Using the linear in $m_\pi$ fit, which seems further justified by the BMW and preliminary super-fine ($a\sim 0.06$~fm) MILC results, returns a value of $\s_{\pi N}\sim67\pm2$~MeV, which is in good agreement with the determination of the ETM Collaboration~\cite{Alexandrou:2008tn}.

As a final note, I want to mention that all the heavy baryon $\chi$PT analyses of nucleon mass results have required one to input the value of the axial coupling, generally with the value $g_A\sim1.2$.  Although possibly acceptable, this is distasteful.  One would really like to see the value of this coupling emerge from a fit to the nucleon mass itself, which would be a smoking gun in a hunt for predicted chiral non-analytic dependence upon the light quark masses.  However, given the state of the lattice results for the nucleon mass, see Fig.~\ref{fig:MNnf2+1}, this seems quite unlikely to happen.  As an alternative, one should move in the direction of global fits, which include both the the nucleon mass and the nucleon axial charge, as well as other observables which can help to constrain the phenomenologically more interesting couplings in the nucleon Lagrangian.

%Thank you!

\begin{acknowledgments}
I would like to acknowledge all the people (Collaborations) who have provided me with their often preliminary results for the nucleon mass.  In particular, I would like to thank Tom Blum, Shoji Hashimoto, Christian Hoelbling, Yoshinobu Kurumashi, Hiroshi Ooki, Gerrit Schierholz and Doug Toussaint.  I would like to thank all the members of the LHP whom I collaborated with on much of this work.  I would like to thank Brian Tiburzi for many lengthy discussions on heavy baryon $\chi$PT and I would like to thank Massimiliano Procura for helping me understand covariant baryon $\chi$PT.  And of course I would like to thank the organizers of \textit{Lattice 2008} for inviting me to give this talk!  This work was supported in part by the U.S. DOE OJI
grant DE-FG02-07ER41527 and DOE grant DE-FG02-93ER40762.
\end{acknowledgments}

\end{document}